\documentclass[journal]{IEEEtran}
\ifCLASSINFOpdf
  \usepackage[pdftex]{graphicx}
\else
\fi
\usepackage{amsmath}
\usepackage{amssymb, amsfonts}

\usepackage{algorithmic}

\usepackage{hyperref}

\usepackage[style=ieee, maxbibnames=6, minbibnames=1]{biblatex}

\def\BibTeX{{\rm B\kern-.05em{\sc i\kern-.025em b}\kern-.08em
    T\kern-.1667em\lower.7ex\hbox{E}\kern-.125emX}}

\begin{document}
%
\title{Deep Learning for Retrospective Motion Correction in MRI: A Comprehensive Review}
%
%
%

\author{Veronika Spieker*, Hannah Eichhorn*, Kerstin Hammernik, Daniel Rueckert, \IEEEmembership{Fellow, IEEE}, Christine Preibisch, Dimitrios C. Karampinos, and Julia A. Schnabel, \IEEEmembership{Fellow, IEEE}
\thanks{*Veronika Spieker and Hannah Eichhorn contributed equally to this work. V.S. and H.E. were supported in part by the Helmholtz Association under the joint research school ”Munich School for Data Science - MUDS”. Correspondence to Julia A. Schnabel: julia@ieee.org.}
\thanks{Veronika Spieker, Hannah Eichhorn, Kerstin Hammernik and Julia A. Schnabel are with the Institute of Machine Learning for Biomedical Imaging, Helmholtz Munich, Germany.}
\thanks{Veronika Spieker, Hannah Eichhorn, Kerstin Hammernik, Daniel Rueckert and Julia A. Schnabel are with School of Computation, Information and Technology, Technical University of Munich, Germany  (e-mail: v.spieker@tum.de, hannah.eichhorn@tum.de, khammernik@tum.de, daniel.rueckert@tum.de, julia.schnabel@tum.de).}
\thanks{Kerstin Hammernik and Daniel Rueckert are also with the Department of Computing, Imperial College London, United Kingdom.}
\thanks{Daniel Rueckert is also with Artificial Intelligence in Healthcare and Medicine, Klinikum rechts der Isar, Technical University of Munich, Germany.}
\thanks{Julia A. Schnabel is also with School of Biomedical Imaging and Imaging Sciences, King’s College London, United Kingdom.}
\thanks{Christine Preibisch is with the Department of Neuroradiology, Klinikum rechts der Isar, Technical University of Munich, Germany (e-mail: preibisch@tum.de).}
\thanks{Dimitrios C. Karampinos is with the Department of Diagnostic and Interventional Radiology,
Klinikum rechts der Isar, Technical University of Munich, 
Germany (e-mail: dimitrios.karampinos@tum.de).}
}

\markboth{}%
{Spieker and Eichhorn \MakeLowercase{\textit{et al.}}: Deep Learning for Retrospective Motion Correction in MRI: A Comprehensive Review}
%



\maketitle

\begin{abstract}
Motion represents one of the major challenges in magnetic resonance imaging (MRI). Since the MR signal is acquired in frequency space, any motion of the imaged object leads to complex artefacts in the reconstructed image in addition to other MR imaging artefacts. Deep learning has been frequently proposed for motion correction at several stages of the reconstruction process. The wide range of MR acquisition sequences, anatomies and pathologies of interest, and motion patterns (rigid vs. deformable and random vs. regular) makes a comprehensive solution unlikely. To facilitate the transfer of ideas between different applications, this review provides a detailed overview of proposed methods for learning-based motion correction in MRI together with their common challenges and potentials. This review identifies differences and synergies in underlying data usage, architectures, training and evaluation strategies. We critically discuss general trends and outline future directions, with the aim to enhance interaction between different application areas and research fields.

\end{abstract}



\section{Introduction}

Motion remains a major challenge for fully exploiting the diagnostic potential of magnetic resonance imaging (MRI). Whereas MRI stands out as a non-invasive medical imaging modality with excellent soft tissue contrast, its intrinsically long acquisition times make it more susceptible to motion than most other modalities. 
However, with the fast development of deep learning in recent years,  many learning-based \textit{motion correction} (MoCo) methods have been proposed to tackle this challenge in a retrospective, data-driven manner. 

Despite the existence of reviews on motion artefacts and classical MoCo \cite{Godenschweger_2016, Zaitsev_2015}, no comprehensive overview of learning-based methods for MR motion correction exists so far. Especially an overview of the increasingly popular field of combined MoCo and image reconstruction is missing, which could foster the transfer of deep learning models between applications. Whereas differences in region of interests, acquisition schemes and motion types intrinsically affect data-driven approaches, synergies in underlying models and overall methods need to be identified.
In this review, we highlight such differences and synergies at all stages of learning-based motion correction by analysing data usage, architectures, training and evaluation strategies. 
Furthermore, we intend to generate a general understanding of recent learning-based MoCo approaches in MRI by outlining respective obstacles and potentials and aim to enhance interaction between the fields of machine learning and MRI. 

We review published articles that present methodological contributions for learning-based retrospective MoCo in MRI. We searched for articles on PubMed and GoogleScholar until August~2023, using combinations of the keywords "\textit{Motion Correction}", "\textit{Motion Compensation}", "\textit{Deep Learning}" and "\textit{Magnetic Resonance Imaging}", from which we selected the most relevant ones.

\noindent The remaining review is structured as follows: 
\begin{itemize}
    \item[\ref{sec:backgr}] \textit{Background}: MR motion artifacts and classical MoCo
    \item[\ref{sec:data}] \textit{Data Availability \& Motion Simulation}: Common brain and cardiac\slash abdominal data strategies
    \item[\ref{sec:arch}] \textit{Architectures}: Image- \& k-space-based MoCo methods
    \item[\ref{sec:obj}] \textit{Training Objectives}: Training strategies and losses
    \item[\ref{sec:eval}] \textit{Evaluation Metrics}: Image Quality, Motion Detection \& Estimation and Downstream Tasks
    \item[\ref{sec:discussion}] \textit{Discussion} of sections \ref{sec:data}, \ref{sec:arch}, \ref{sec:obj} and \ref{sec:eval} 
    \item[\ref{sec:outlook}] \textit{Conclusion \& Outlook}.
\end{itemize}

\section{Background}
\label{sec:backgr}
\begin{figure*}[!t]
\includegraphics[width=\textwidth]{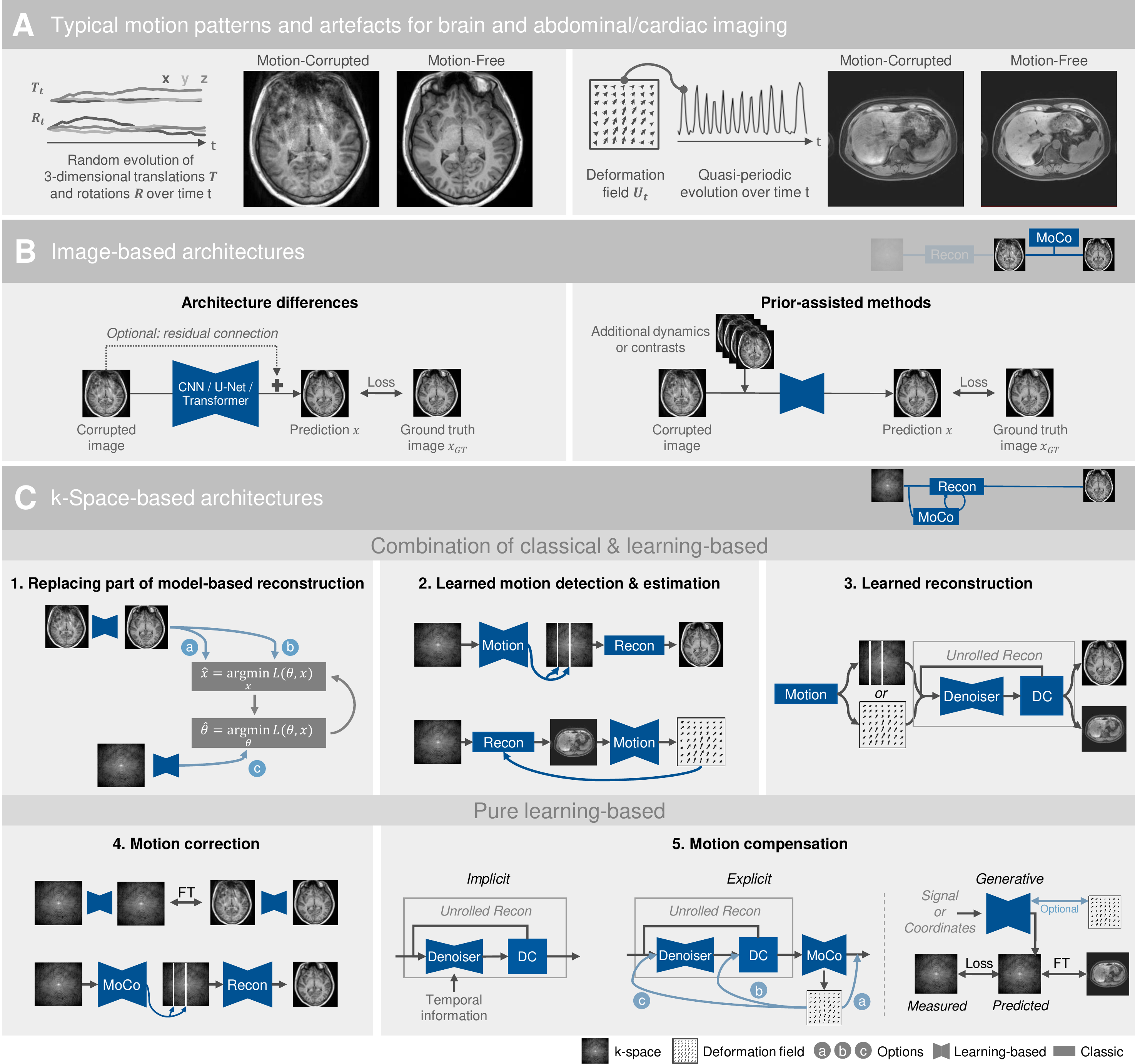}
\centering
\caption{\textbf{(A)} Typical motion patterns for brain and abdominal/cardiac imaging, together with examples of motion-corrupted and motion-free images (brain images from \cite{Ganz_2022}, remaining data acquired at Klinikum Rechts der Isar, Munich). For brain imaging, random rigid-body motion is typically assumed, which results in blurring and ringing artefacts, depending on the exact acquisition scheme and motion pattern. For cardiac and abdominal imaging, motion is typically deformable and quasi-periodic, leading to blurring and possibly, ghosting artefacts. \textbf{(B)} Visualisation of image-based MoCo with motion-corrupted images as input and motion-corrected images as output of a neural network (CNN, U-Net or transformer). Optionally, a residual connection enforces the network to learn artefact maps. Prior-assisted methods incorporate additional information, like additional dynamics or contrasts. \textbf{(C)} Illustration of k-space-based MoCo. Methods that combine classical and learning based modules can be categorized as follows: \textit{(1)} methods that replace different components of a model-based reconstruction, which iterates between finding an image $x$ and corresponding motion parameters $\theta$ by minimizing a loss function, $L(\theta,x)$. Learning-based modules target (a) the image initialization, (b) the loss function or (c) the motion parameters. 
\textit{(2)} Combining a classical reconstruction with a learning-based motion detection of corrupted k-space measurements or a learning-based estimation of motion fields, $\mathbf{U_t}$. \textit{(3)} Combining classical motion detection or estimation with a learned unrolled reconstruction that iterates between denoising networks and data consistency (DC) blocks. Purely learning-based methods include \textit{(4)} approaches that directly aim to correct motion artefacts by either performing convolutions in k-space and image-space or excluding motion-corrupted k-space measurements from the learning-based reconstruction and \textit{(5)} motion compensation methods which use motion information implicitly or explicitly to achieve higher quality image reconstructions. A subclass of approaches fit a generative reconstruction model to raw data of an individual, whereas motion can optionally be modelled explicitly by including deformation fields.  
For all visualizations in \textbf{(B)} and \textbf{(C)} we illustrated the most prevalent anatomy, even though the anatomies are interchangeable in most cases.}
\label{fig:graph_abstr}
\end{figure*}
As a basis for understanding learning-based MoCo approaches, we summarize the fundamental principles of MR motion artefacts and classical motion correction in the following section. Please refer to \cite{Godenschweger_2016, Zaitsev_2015} for more detailed overviews.

Relevant motion during MR image acquisition comprises both moving organs - e.g. due to cardiac motion or respiration - and conscious or unconscious movement of body parts - e.g. due to patient discomfort. Different motion patterns can be observed when imaging different body regions: For brain imaging, the movement of the head is usually assumed to be random and characterized by six rigid-body motion parameters (three rotational and three translational components), commonly neglecting small deformable motion, e.g. due to brain pulsation. In contrast, for abdominal and cardiac imaging the intrinsic movement of organs due to breathing and heartbeat leads to quasi-periodic patterns and deformable, non-rigid motion with significantly more degrees of freedom. For fetal body imaging, 
next to quasi-periodic motion from the fetus and mother, further vast and unpredictable sudden (non-)rigid motion may occur due to sudden movement of the fetus. 
Regardless of the exact pattern, motion of the imaged object affects the MR signal, which is acquired in frequency space (or k-space). On the one hand, changes in position disrupt the capability to encode spatial information in the acquired signal. On the other hand, physical MR signal properties are negatively influenced by second-order motion effects, e.g. due to motion-induced magnetic field inhomogeneities or spin history effects. Thus, after reconstructing motion-corrupted data from frequency to image space, complex artefacts may arise, which cannot be corrected in a straightforward process \cite{Godenschweger_2016}. 
Exemplary motion-corrupted images for brain and abdominal imaging in Fig.~\ref{fig:graph_abstr}A illustrate that motion may hinder successful diagnoses. The versatility of MRI protocols and motion types
makes a comprehensive solution unlikely.

Several strategies have been proposed to mitigate motion artefacts. First, subject motion can be constrained physically, for instance by acquiring abdominal scans only during breath hold \cite{McClelland_2013} or using sedation or general anaesthesia when imaging young children \cite{Chavhan_2022, Uffman_2017}. 
Second, image acquisition schemes have been designed to be more robust towards motion, either by selectively acquiring data in certain motion states or using advanced sampling patterns \cite{Pipe_1999, Winkelmann_2007, Peters_2000}. 
Third, accelerated and parallel imaging methods have been introduced, which offer the advantage of shorter acquisition times corresponding to less opportunity for motion events~\cite{Pruessmann_1999, Lustig_2007}.

Next to these mitigation strategies, which are still susceptible to motion artefacts, another group of approaches have been proposed to directly perform \textit{motion correction} by explicitly removing motion artefacts, and \textit{motion compensation} by leveraging the regularity of motion patterns for a better reconstruction.
These approaches include prospective methods, which are applied during image acquisition \cite{Zaitsev_2006, vanderKouwe_2006, Bush_2021}, and retrospective methods, which are applied after image acquisition at various points in the reconstruction pipeline \cite{Atkinson_1997, CorderoGrande_2016}. Retrospective methods must cope with motion-induced image information loss, e.g. due to data inconsistencies in k-space \cite{Zaitsev_2015}. For this, deep learning methods are particularly promising due to their capability to identify complex patterns in the absence of a complete analytical model. 

Note that the following sections focus on motion correction and compensation of motion artefacts that originate in the acquisition process, i.e. the k-space domain, and reconstruction refers to the domain-transfer from k-space to image-space. For strategies targeting slice-to-volume reconstruction (SVR) of highly accelerated (and hence nearly artefact-free) 2D slices, such as dominantly applied for fetal motion correction, we refer the reader to \cite{Uus_2023}. 



\section{Data Availability and Motion Simulation} \label{sec:data}
The majority of learning-based MoCo methods rely on supervised training and thus, on the availability of paired data with and without motion artefacts. Even unsupervised or self-supervised approaches use paired data for quantitative performance evaluations (compare Section \ref{sec:eval}). Some authors acquire pairs of motion-corrupted and ground truth (GT) motion-free images for training and evaluation. However, it is costly and not always feasible to acquire large paired datasets, which is why motion simulations are commonly used.

When simulating motion artefacts, it is important to consider the typical motion patterns of the anatomy of interest, which we described in Section~\ref{sec:backgr}.
In the following, we summarize the common simulation procedures for brain as well as cardiac and abdominal imaging.

\subsection{Brain}
\label{sec:data_brain}
The simulation of rigid-body motion follows the MRI forward model in the presence of motion \cite{Atkinson_2022}:
\begin{equation}
    y = \sum_{t=1}^{T} \mathbf{M}_t \mathcal{F} \mathbf{U}_t x,
    \label{eq:motion_forward_model}
\end{equation}
where the Fourier transform $\mathcal{F}$, the sampling mask $\mathbf{M}_t$ and the motion transform $\mathbf{U}_t$, are applied to the GT image $x$ for each time point $t$ to generate the motion-corrupted k-space $y$. In the case of rigid-body motion, the motion transform, $\mathbf{U}_t = \mathbf{T}_t\mathbf{R}_t$, consists of rotation and translation transforms, $\mathbf{R}_t$ and $\mathbf{T}_t$.
Additionally, coil sensitivities or second-order motion effects can be included in the forward model to extend the simulation to the specific application.

It is mathematically equivalent to simulate motion in image space or in k-space \cite{Loktyushin_2013}. 
\begin{figure}[!t]
\includegraphics[width=\columnwidth]{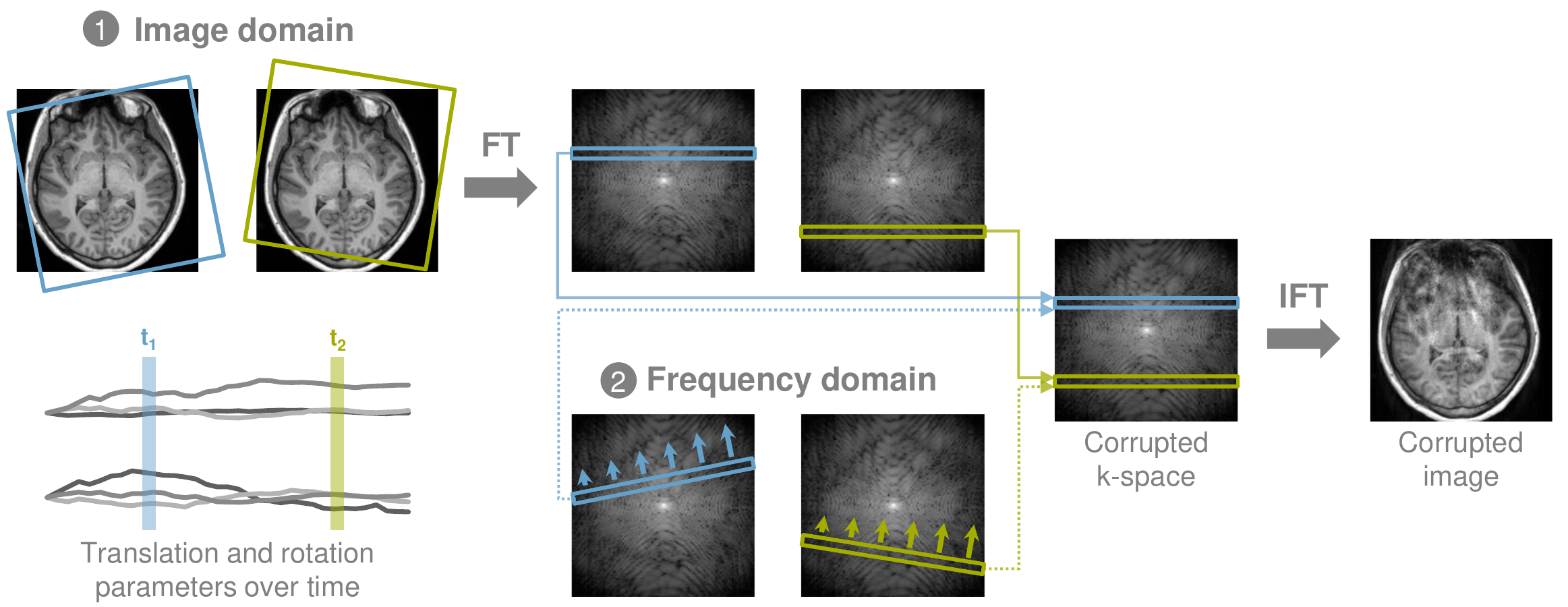}
\centering
\caption{Rigid-body motion simulation in image or frequency domain, based on x-, y- and z-translation and rotation parameters for each time step. When simulating in image domain (1), translation and rotation parameters are applied to the image. When simulating in frequency domain (2), corresponding k-space lines are rotated and multiplied with linear phase ramps (visualized by arrows). For both, k-space lines of different time steps are merged into one corrupted k-space.}
\label{fig:simulation}
\end{figure}
As visualized in Fig.~\ref{fig:simulation}, simulations in image space are performed by rotating and translating the image and replacing the corresponding k-space lines with the Fourier transform of the transformed image for each time step. Simulations in k-space are based on the properties of the Fourier transform: rotations of the imaged objects correspond to equivalent rotations in k-space and translations $\mathbf{T}$ correspond to multiplications with linear phase ramps depending on the translation parameter $a$ and k-space coordinate in readout direction $k_{RO}$:
\begin{equation}
    \mathbf{T} (y) =  y \cdot \exp(-i2\pi a k_{RO}).
    \label{eq:motion_kspace}
\end{equation}
 Regardless of the domain, in which the simulations are performed, it is important to match the timing of the motion to the MR acquisition scheme to simulate realistic artefacts \cite{Levac_2022b}.

\subsection{Cardiac and Abdominal}
\label{sec:data_body}
Following \eqref{eq:motion_forward_model}, image-based non-rigid motion simulation is achieved by applying a deformable vector field (DVF) as motion transform $\mathbf{U}_t$. Realistic DVF can be obtained by registering reference images, e.g. from different motion states. Statistical modulation of the DVF allows augmentation for training purposes. If available, multiple time-resolved reconstructions $x_t$ can be used to substitute $\mathbf{U}_t x$ and, hence, simulate without DVF. 



In contrast to rigid motion, it is not trivial to simulate non-rigid deformation in k-space directly. Therefore, cardiac and respiratory motion simulation can be approximated with varying translations in k-space. To simulate periodic motion, linear phase ramps with a periodically varying translation parameter, i.e.  $a(t) \propto \sin(t) $, are applied in \eqref{eq:motion_kspace}. It needs to be noted that this is a strongly simplified motion representation.

\section{Architectures} \label{sec:arch}
This section covers an overview of proposed architectures for learning-based MoCo in MRI. Architectures can be categorized into (A) image-based and (B) k-space-based. Each section includes methods applied to different anatomies and highlights similarities, differences and general trends.
\subsection{Image-Based Motion Correction}
Image-based MoCo methods take motion-affected images as input and produce motion-corrected images as output, similar to image denoising or deblurring tasks, as sketched in Fig.~\ref{fig:graph_abstr}B. They differ based on their (a) underlying network architecture and (b) potential use of prior information. 

\paragraph{Underlying Network Architectures}
The image-to-image translation task can be solved using convolutional neural networks (CNNs), consisting of several consecutive convolutional layers with corresponding activation functions \cite{Sommer_2020,Tamada_2020,Chatterjee_2020,Pirkl_2022}. Some approaches process the input images in multiple resolutions, using varying patch and kernel sizes~\cite{Sommer_2020} or dilated convolutions~\cite{Pirkl_2022}.
Similarly, convolutional encoder-decoder structures can be utilized, which consist of a downsampling path for feature encoding, followed by an upsampling path for decoding the extracted features~\cite{Pawar_2020,Liu_2020,Chatterjee_2020,Lee_2021,Oksuz_2021,Duffy_2021,Xu_2022,Jafari_2022,AlMasni_2022,AlMasni_2023,Johnson_2019,Usman_2020,Bao_2022,Kustner_2019,Wu_2021,Ghodrati_2021b,Liu_2021,Ghodrati_2021,Oh_2021}. 
As common for U-Nets, most methods use skip connections to transfer details from the encoder to the decoder \cite{Pawar_2020,Liu_2020,Chatterjee_2020,Oksuz_2021,Duffy_2021,Xu_2022,AlMasni_2022,Johnson_2019,Usman_2020,Bao_2022,Kustner_2019,Wu_2021,Ghodrati_2021b,Liu_2021,Oh_2021}. Additionally, some approaches cascade multiple encoder-decoder structures consecutively \cite{Liu_2020,AlMasni_2022,Kustner_2019,Wu_2021}. 
Further, to exploit temporal dependencies of dynamic data, a recurrent encoder-decoder network has been proposed, replacing convolutions with convolutional long/short-term memory models \cite{Abdi_2023}.
Inspired by the recent success of transformer architectures in other fields, initial works employ self-attention mechanisms to address long-distance spatial dependencies of motion artefacts \cite{vanderGoten_2022,Armanious_2020}.

The above presented approaches vary with respect to whether the inputs of the networks are patches \cite{Sommer_2020,Tamada_2020,Pirkl_2022,Liu_2020,Duffy_2021,Johnson_2019,Ghodrati_2021b,Ghodrati_2021} or complete images \cite{Chatterjee_2020,Pawar_2020,Chatterjee_2020,Lee_2021,Oksuz_2021,Xu_2022,Jafari_2022,AlMasni_2022,AlMasni_2023,Usman_2020,Kustner_2019,Wu_2021,Liu_2021,Oh_2021,Armanious_2020,vanderGoten_2022,Abdi_2023}. The networks are either implemented as classical \cite{Chatterjee_2020,Pawar_2020,Chatterjee_2020,Lee_2021,Oksuz_2021,Duffy_2021,Xu_2022,Jafari_2022,AlMasni_2022,AlMasni_2023,Johnson_2019,Usman_2020,Kustner_2019,Wu_2021,Ghodrati_2021b,Liu_2021,Ghodrati_2021,Armanious_2020,vanderGoten_2022,Abdi_2023} or residual networks \cite{Sommer_2020,Tamada_2020,Pirkl_2022,Liu_2020,Bao_2022,Oh_2021}, generating corrected images or motion artefact maps, respectively. 
To further improve the performance, the MoCo task can be combined with learning-based downstream tasks in an end-to-end manner, where both tasks can benefit from each other \cite{Xu_2022}. 

\paragraph{Prior-Assisted Methods}
The presented architectures can be modified to take advantage of additional information, like different contrasts \cite{Chatterjee_2020,Lee_2021}, multi-echo or multi-parametric acquisitions \cite{Pirkl_2022,AlMasni_2023}, similar slices \cite{Chatterjee_2020,AlMasni_2022} or dynamic information \cite{Tamada_2020,Lyu_2021,Jafari_2022,Ghodrati_2021b}. 
These prior-assisted methods process multiple inputs by either multiple or shared encoders and decoders with shared feature extraction \cite{Chatterjee_2020,Lee_2021,AlMasni_2023}, by concatenating the inputs on different channels \cite{Chatterjee_2020,Tamada_2020,Pirkl_2022,Jafari_2022,AlMasni_2022} 
or by using a recurrent network structure for the additional dimension \cite{Lyu_2021}. For instance, Ghodrati \emph{et al.} \cite{Ghodrati_2021b} attempt to leverage temporal information by computing a loss on the features extracted by an auxiliary network pretrained on dynamic images.
Moreover, dynamic information can be utilized in registration-based methods, where a CNN is used to register binned data into a common space, the combination of which results in a motion-corrected output image \cite{Lv_2018}.

\subsection{k-Space-Based Motion Correction}
\label{sec:k_space}
Contrary to image-based methods, MoCo can also leverage the additional information content of raw k-space data and thus, interact with the MR reconstruction process (see Fig.~\ref{fig:graph_abstr}C). Different components of the motion-aware reconstruction pipeline can be learning-based. In the following sections we provide an overview of methods which combine classical and learning-based modules, and methods with pure learning-based modules.

\subsubsection{Combination of Classical and Learning-Based Approaches} \label{sec:arch_k_comb}
Multiple methods extend classical frameworks with individual learning-based MoCo or reconstruction components. A part of a model-based reconstruction, the motion analysis or the reconstruction itself can be learned.

\paragraph{Replacing Part of Model-Based Reconstructions} \label{par:repl_model_based}
Model-based MoCo algorithms rely on the joint estimation of motion parameters and the reconstructed image. Various approaches propose to replace different parts of these optimisation procedures with learning-based components to enable faster convergence and ideally, more stable reconstructions (Fig.~\ref{fig:graph_abstr}C.1). 
Kuzmina \emph{et al.}~\cite{Kuzmina_2022} use a CNN as part of the loss function for autofocusing, where the optimisation is based on an image quality metric. For data consistency (DC)-based optimisation procedures, CNNs or U-Nets are employed for motion parameter estimation \cite{Hossbach_2022}, as initialisation of the motion-corrected image \cite{Haskell_2019,Wang_2020} or as reconstruction networks whose weights are defined by a hypernetwork that is dependent on the motion parameters \cite{Singh_2023}.  
In contrast, Levac \emph{et al.}~\cite{Levac_2022} propose an unsupervised approach, using a score-based model, that was trained on motion-free images, in the joint estimation of image and motion parameters.
All these approaches have in common that the network is pre-trained and used as plug-and-play component during test-time optimisation. 
Moreover, all these approaches focus on rigid-body motion, having considerably less degrees of freedom than non-rigid motion. 

\paragraph{Learning-Based Motion Analysis and Classical Reconstruction} \label{par:learning_motion_class_recon}
Another group of methods leverage the random nature of rigid-body motion. As visualized in Fig.~\ref{fig:graph_abstr}C.2, they learn detection models for motion-affected k-space measurements and inform classical reconstruction procedures with the extracted motion timing. Eichhorn \emph{et al.}~\cite{Eichhorn_2023} employ a CNN for a line detection in k-space and use these line-wise classification labels as weights in the DC term of a total variation-based reconstruction procedure. Cui \emph{et al.}~\cite{Cui_2023} train an image-based CNN to correct motion artefacts and compare the k-space of the original and motion-corrected images to generate undersampling masks for motion-affected k-space lines. Undersampled original data are then reconstructed with a classical compressed sensing procedure. 

In the case of quasi-periodic motion, the assumption of individual motion-corrupted k-space lines does not apply. Rather than correcting for single motion events, motion compensation methods leverage the periodicity of motion for higher-quality reconstructions of undersampled data. These methods learn motion estimates, which are included in a model-based classical reconstruction \cite{Qi_2021, Munoz_2022, Hammernik_2021, Pan_2022, Feinler_2023}. Motion fields are predicted using image-based registration and integrated into the forward operator of the reconstruction problem. Existing approaches vary regarding the registration network's input, i.e. complete image \cite{Hammernik_2021, Pan_2022, Feinler_2023} vs. image patches \cite{Qi_2021, Munoz_2022}, and paired \cite{Qi_2021, Munoz_2022} vs. grouped input \cite{Hammernik_2021, Pan_2022, Feinler_2023}. Furthermore, the motion estimation network can be pre-trained \cite{Qi_2021, Munoz_2022, Hammernik_2021} or optimized jointly with the reconstruction problem \cite{Pan_2022}. A hybrid approach is proposed in \cite{Feinler_2023}, where a motion estimation is obtained with a pre-trained multi-scale network, and consecutively optimized in an iterative reconstruction. Munoz \emph{et al.} \cite{Munoz_2022} leverage a diffeomorphic registration network to predict forward and backward motion fields in one run rather than individually.
A different application of learned motion estimates is presented in \cite{Shao_2022}, where real-time high-quality reconstructions are obtained by deforming a reference image with motion fields predicted from few k-space lines. Whereas registration is conducted in image-space, classical reconstruction layers are used to calculate the loss in k-space.

\paragraph{Classical Motion Analysis and Learning-Based Reconstruction}
\label{par:class_motion_learning_recon}
In contrast, motion detection and estimation can also be performed classically and combined with a learned unrolled DC-based reconstruction (Fig.~\ref{fig:graph_abstr}C.3). Rotman \emph{et al.}~\cite{Rotman_2021} detect discrete motion timings by comparing signals from two opposite coil elements and learn an unrolled reconstruction, in which the regularising network separately receives the data acquired in the dominant and remaining motion states. 
Miller \emph{et al.}~\cite{Miller_2023} employ a classical spatio-temporally constrained registration of dynamic images to a single motion state. The registered images are encoded and forwarded into an unrolled reconstruction, which is trained in a self-supervised manner by splitting the available data into subsets.

\subsubsection{Pure Learning-Based Approaches} \label{sec:arch_k_pure_dl}
Compared to the previous section, several methods combine MoCo and image reconstruction in one purely learning-based framework. These can be distinguished by their aim to either correct or compensate for motion.
%
%
\paragraph{Motion Correction}
Proposed motion correction approaches explicitly aim to remove motion artefacts in the underlying data (Fig.~\ref{fig:graph_abstr}C.4). Singh \emph{et al.}~\cite{Singh_2022} propose a network consisting of interleaved or alternating convolutions in image and k-space for simultaneous rigid-body MoCo and reconstruction. This approach was further developed into a data consistent method, which we already introduced in section \ref{par:repl_model_based} \cite{Singh_2023}. Oksuz \emph{et al.}~\cite{Oksuz_2019}  realize data consistent reconstructions of cardiac data with ECG mistriggering artefacts in line with the methods presented in section \ref{par:learning_motion_class_recon}. They propose to employ a CNN to learn undersampling masks for motion-affected k-space lines and reconstruct the undersampled data with a recurrent network. 
In an extension, they train the detection and reconstruction networks end-to-end with a segmentation network and thus, optimize the MoCo specifically for the downstream task of interest \cite{Oksuz_2020}.

\paragraph{Motion Compensation}
Presented motion compensation methods leverage occurring motion to improve reconstruction results along with accelerated acquisition times, as illustrated in Fig.~\ref{fig:graph_abstr}C.5. This can be achieved \textit{implicitly} by including the temporal dimension in the denoising process of an unrolled reconstruction \cite{Schlemper_2018, Kustner_2020, Terpstra_2022,Qin_2019}. Spatial and temporal convolutions are applied to dynamic image series either with a joint \cite{Schlemper_2018} or separated spatio-temporal kernel, in a cascaded \cite{Kustner_2020} or parallel manner \cite{Terpstra_2022}. To leverage further information from adjacent frames, Schlemper \emph{et al.} \cite{Schlemper_2018} include a data sharing layer. Terpstra \emph{et al.} \cite{Terpstra_2022} extend the implicit motion-compensated reconstruction with motion fields obtained from a pretrained model. Qin \emph{et al.} \cite{Qin_2019} employ recurrent networks to exploit dependencies along temporal dimensions as well as along stages of the iterative reconstruction.

In contrast, several methods \textit{explicitly} learn the motion model with the reconstruction problem in an end-to-end fashion \cite{Huang_2021, Seegoolam_2019, Yang_2022, Qi_2021b, Kustner_2022, Gan_2022}. Huang \emph{et al.} \cite{Huang_2021} append motion estimation and correction modules to a reconstruction network and train the framework with one combined loss function. Others directly feed learned motion estimates into the unrolled reconstruction process, either as input of the denoiser \cite{Seegoolam_2019, Yang_2022} or in the DC layer \cite{Qi_2021b, Kustner_2022}. Additionally, these methods differ in the way motion estimates were obtained, e.g. using optical flow \cite{Seegoolam_2019}, groupwise registration \cite{Yang_2022}, patch-wise registration \cite{Qi_2021b} or registration in k-space \cite{Kustner_2022}. A different approach is presented by Gan \emph{et al.} \cite{Gan_2022}, where a motion estimation network is leveraged to train a reconstruction framework in an unsupervised manner, i.e. by deforming other dynamics for loss calculation. Whereas motion is modelled explicitly during training, it is implicitly represented in the reconstruction network at inference.

Whereas all previously presented methods have been developed for subject-independent inference, a few \textit{generative} motion-aware reconstruction methods train a reconstruction model per subject to infer for that same individual \cite{Zou_2022, Yoo_2021, Huang_2022, Feng_2022, Catalan_2023, Spieker_2023}. Due to their distinct training strategy, we consider these methods as a separate category. Still, motion modeling can be implicit and explicit. In particular, quasi-periodic motion can be modeled as a latent manifold and then transformed into dynamic images through a more complex representation \cite{Yoo_2021}. Transformation of the resulting images into Fourier space allows for network optimization in a self-supervised manner. A different motion modelling strategy \cite{Zou_2022} learns a low-dimensional signal, which is mapped to motion estimates. These are then applied to one learned reference reconstruction. Again, predicted images are compared with the acquired data points in Fourier space. 
Recently, implicit neural representations (INR) have also gained attention for dynamic MR reconstruction. Based on spatial and temporal coordinates, a light-weight network predicts the corresponding intensity values in image- \cite{Feng_2022, Catalan_2023} or k-space \cite{Huang_2022, Spieker_2023}. By including the cardiac \cite{Feng_2022, Catalan_2023, Huang_2022} or respiratory phase \cite{Spieker_2023} as temporal dimension, motion is implicitly modelled and motion-resolved reconstructions can be obtained at inference. Whereas k-space-based INRs \cite{Huang_2022, Spieker_2023} can directly be compared with the acquired points, image-based INRs require transformation to k-space, either by applying the non-uniform fast Fourier transform on a fully queried image \cite{Feng_2022} or taking advantage of the Fourier Slice Theorem for individual spokes \cite{Catalan_2023}. To enable a better spectral representation the proposed approaches apply Fourier \cite{Huang_2021, Spieker_2023}, spatiotemporal Fourier \cite{Catalan_2023} or Hash encoding \cite{Feng_2022}. 

\section{Training Objectives} \label{sec:obj}

 \subsection{Image-Based Motion Correction}
\begin{figure}
\includegraphics[width=\columnwidth]{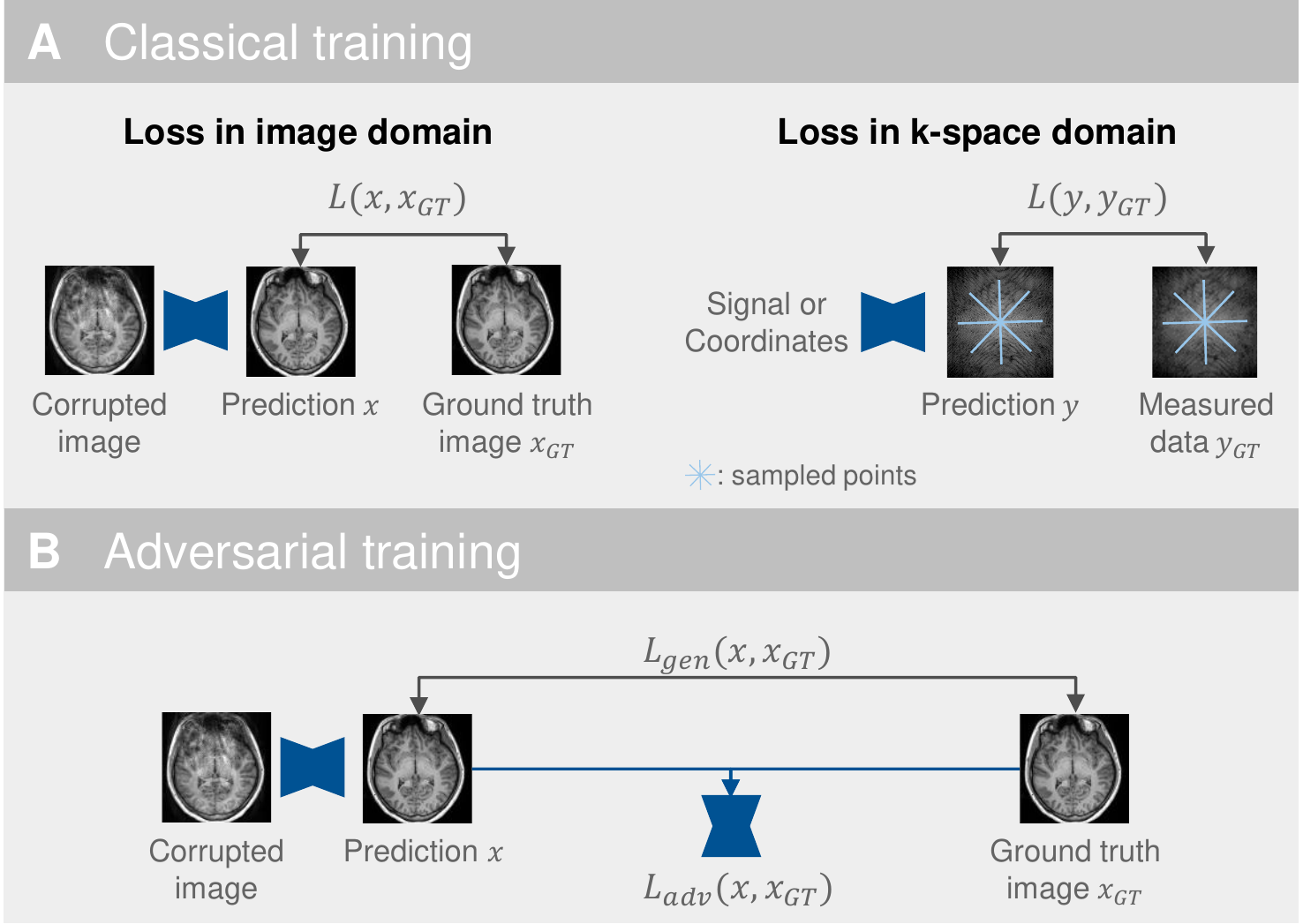}
\centering
\caption{Visualisation of classical and adversarial training objectives for learning-based MoCo. (A) For classical training, the loss can be calculated in image or k-space domain. In image domain, the predicted image is compared with the ground truth image, e.g. using a voxel intensity-based loss function $L$. In k-space domain, predicted k-space values are compared with the measured data at the sampled locations. (B) For adversarial training, next to the generative loss $L_{gen}$, an additional discriminator network is trained to compete with the generator network and distinguish the predicted image from the ground truth ($L_{adv}$).}
\label{fig:objectives}
\end{figure}

As visualized in Fig.~\ref{fig:objectives}A (left), classical network training of image-based MoCo methods in a supervised setting is performed by calculating a voxel intensity-based cost function between the network's prediction and a ground-truth motion-free image. Typical intensity-based cost functions are the L1 and L2 loss (which stand for the mean absolute and mean squared error, respectively) or the structural similarity index \cite{Wang_2004}. Please refer to Sec. \ref{sec:eval_img_qu} for mathematical definitions. Next to these examples, any other image similarity metric can be used as cost function. 

A different training objective, however, is employed with conditional generative adversarial networks (GANs), as illustrated in Fig.~\ref{fig:objectives}B. 
A generator network, mapping the motion-corrupted to a motion-free image, is extended with a discriminator network, which aims to distinguish the predicted image from a ground truth image. Several supervised GAN-based methods have been proposed for various anatomies \cite{Johnson_2019, Usman_2020, Bao_2022, Kustner_2019, Wu_2021, Ghodrati_2021b}. 
Next to the adversarial loss, some of these methods rely on voxel intensity-based cost functions as generator loss to compare the predicted and ground truth image \cite{Johnson_2019, Usman_2020}. Others include a perceptual loss \cite{Bao_2022, Kustner_2019, Wu_2021}, style transfer loss \cite{Kustner_2019} or structural similarity loss (SSIM) \cite{Ghodrati_2021b} to account for global changes. Bao \emph{et al.} \cite{Bao_2022} propose an additional entropy loss to enhance image homogeneity. 
Next to the adversarial approach, Küstner \emph{et al.} \cite{Kustner_2019} present another supervised generative training strategy using a variational autoencoder (VAE), which attempts to learn a motion-free latent distribution directly from the image pair.
%
%

To cope with the lack of paired motion-free and corrupted data, unsupervised generative models aim to correct for motion from unpaired data \cite{Armanious_2020, Liu_2021, Ghodrati_2021,Oh_2021}. The CycleGAN architecture consisting of two GANs is adapted in~\cite{Liu_2021} and~\cite{Armanious_2020}. Two generators, one corrupting a motion-free image and one correcting an unpaired corrupted image, are trained to invert each other (cycle-transform). Whereas the adversarial loss can be computed with an unpaired image from the other domain, the generative loss is calculated in an unsupervised manner on the cycle-transformed input from the same domain. Liu \emph{et al.} \cite{Liu_2021} additionally disentangle the latent representation of the generators into artefact and content information, and train the network with images generated from content-swapped translations. Both \cite{Liu_2021} and \cite{Armanious_2020} include multi-scale cost functions. In a different setting, Oh \emph{et al.} \cite{Oh_2021} treat motion as a probabilistic undersampling problem and train a generator to remove undersampling artefacts. They attempt to correct motion-corrupted measurements by combining repeated randomly undersampled reconstructions. 
In contrast to CycleGANs, Ghodrati \emph{et al.} \cite{Ghodrati_2021} regularize the latent space of a single MoCo autoencoder by applying a discriminator with unpaired motion-free images.

\subsection{k-Space-Based Motion Correction}
k-space-based MoCo methods can be trained by comparing the final reconstruction with a ground-truth motion-free image, similarly to image-based methods. 
However, the loss can also be computed in the acquisition domain directly. By comparing the predicted with the available sampled k-space data, MoCo methods can be trained in a self-supervised manner, as visualised in Fig.~\ref{fig:objectives}A (right). The L2 loss is frequently adapted for comparison of the predicted and measured k-space values \cite{Miller_2023, Zou_2022, Yoo_2021}. Due to the inherent nature of higher magnitudes towards the center of k-space, adaptations such as the L2 loss normalized by the square magnitude \cite{Feng_2022} or a high dynamic range loss \cite{Huang_2022} have been proposed, thereby allowing for a more balanced weighting of low and high-frequency components. The k-space based loss can be extended with any further image-based constraint, such as temporal total variation \cite{Feng_2022} to enforce smoothness between dynamics.

Next to the final reconstruction losses, further motion estimation and detection losses can be incorporated into the training objective. MoCo methods including explicit motion modelling can include an image similarity metric on a spatially transformed image in a self-supervised fashion, as well as spatial or temporal smoothness constraints on the predicted motion field \cite{Hammernik_2021, Qi_2021}. Models based on motion detection can be trained with any classification loss reflecting the correct identification of motion-affected lines, such as binary cross entropy \cite{Eichhorn_2023, Oksuz_2020}.


\section{Evaluation Metrics} \label{sec:eval}
Since MoCo is performed as a means to an end for a high-quality image reconstruction, the performance of the presented methods is predominantly evaluated based on their final outcome, using image quality measures. However, some authors also evaluate intermediate motion estimates or make use of downstream tasks. In the following, we provide an overview of the most common evaluation strategies. Whereas we focus on evaluation metrics, most of the presented measures can also be used as loss functions, depending on the architecture and the type of training.

\subsection{Image Quality} \label{sec:eval_img_qu}
Quantitative or qualitative image quality evaluation can be performed by calculating image quality metrics or expert image quality rating, respectively. Quantitative image quality metrics can be either full-reference metrics, which assess the image quality by comparison to a GT reference image, or reference-free metrics, which do not rely on a separate GT image. Due to the variability of motion artefacts, no single image quality measure is sensitive to all possible artefacts.

The majority of the methods presented in section \ref{sec:arch} use two full-reference metrics that attempt to mimick human visual perception: structural similarity index (SSIM) \cite{Wang_2004}, which assesses the degradation of structural information, and peak signal-to-noise ratio (PSNR), which contrasts pixel-wise errors with the maximum signal intensity. Less frequently used full-reference metrics are mean squared error (MSE), root MSE (RMSE), normalized RMSE (NRMSE), mean absolute (percentage) error (MAE/MAPE), normalized mutual information (NMI) \cite{Studholme_1999} and visual information fidelity (VIF) \cite{Sheikh_2006}. Moreover, reference-free metrics like signal-to-noise ratio (SNR) \cite{Dietrich_2007}, contrast-to-noise ratio (CNR) and Tenengrad \cite{Krotkov_1988} are used to to assess image quality without a reference image.
Table \ref{tab:metrics} provides the definitions of these metrics in a consistent notation. Note that across literature there is no standard normalisation constant for NRMSE. Also, all metrics are applied to real-valued images, whereas there is no clear indication on how to handle complex features.
\begin{table}
\caption{Definition of the most commonly used real-valued image quality metrics. }
\label{table}
\setlength{\tabcolsep}{3pt}
\begin{tabular}{|p{65pt}|p{165pt}|}
\multicolumn{2}{p{230pt}}{\textbf{Full-reference metrics}}\\
\hline
 SSIM & 
 $\frac{1}{\lvert \mathcal{M} \rvert}\sum_{m,\hat{m} \in \mathcal{M}}\frac{(2\mu_m\mu_{\hat{m}}+c_1)(2\sigma_{m\hat{m}}+c_2)}{(\mu_m^2 + \mu_{\hat{m}}^2+c_1)(\sigma_m^2 + \sigma_{\hat{m}}^2+c_2)}$ \\[0.3cm]
 MSE & $\frac{1}{IJ}\sum_{i=1,j=1}^{I,J} (x_{ij}-\hat{x}_{ij})^2$ \\[0.3cm]
 PSNR & $10 \log_{10}\frac{\text{max}(\hat{x})^2}{\text{MSE}(x, \hat{x})}$ \\[0.3cm]
 RMSE & $\sqrt{\text{MSE}(x, \hat{x})}$ \\[0.3cm]
 NRMSE &  $\frac{\text{RMSE}(x, \hat{x})}{c}, \text{with varying constant, e.g. }c=\frac{1}{\lvert \mu_{\hat{x}} \rvert}$ \\[0.3cm]
 MAE & $\frac{1}{IJ}\sum_{i=1,j=1}^{I,J} \lvert x_{ij}-\hat{x}_{ij} \rvert$ \\[0.3cm]
 MAPE & $\frac{\text{MAE}(x, \hat{x})}{\text{max}(\hat{x})}$ \\[0.3cm]
 NMI  & $\frac{\text{H}(x)+ \text{H}(\hat{x})}{\text{H}(\mathbf{x,\hat{x}})}, \text{with H}(Z) := -\sum_{z \in Z}z\log z$ \\[0.3cm]
 VIF & \textit{Due to the complexity, we refer the reader to the original publication \cite{Sheikh_2006}}. \\[0.1cm]
\hline
\multicolumn{2}{p{230pt}}{ }\\
\multicolumn{2}{p{230pt}}{\textbf{Reference-free metrics}}\\
\hline
 SNR & $20\log \frac{\mu_s}{\sigma_n}$, \  \\[0.3cm]
 CNR & $20\log \frac{ \lvert \mu_{s1} - \mu_{s2} \rvert}{\sigma_n} $, \ \\[0.3cm]
 Tenengrad & $\frac{1}{IJ}\sum_{i=1,j=1}^{I,J} (\nabla_x x_{ij}^2 + \nabla_y x_{ij}^2)$ \\
\hline 
\multicolumn{2}{p{230pt}}{ }\\
\multicolumn{2}{p{230pt}}{$x$: image to be evaluated, $\hat{x}$: reference image, $m$/$\hat{m}$: patch of $x$/$\hat{x}$, \newline $\mu$: mean value, $\sigma$: standard deviation, $c_1/c_2\propto L^2$: variables proportional to dynamic range $L$, $n$: noise region, $s$: region of interest, $s_1,s_2$: two separate regions in region of interest  $s$, $n$: noise region, $\nabla_{(x,y)}$: gradient in $x$- or $y$-direction}\\
\end{tabular}
\label{tab:metrics}
\end{table}
%

Several approaches also include qualitative image quality evaluation, i.e. through subjective scoring of the reconstructed images by (blinded) experts \cite{Duffy_2021,Kustner_2019,Liu_2020,Pawar_2020,Sommer_2020,Oh_2021,Ghodrati_2021b,Tamada_2020,Jafari_2022,Lv_2018,Ghodrati_2021,Qi_2021b,Kustner_2020,Abdi_2023}. However, there is no standardized way for observer scoring and it varies strongly regarding: 
\begin{itemize}
    \item  evaluation categories and instructions for evaluators (e.g. overall quality, sharpness, diagnostic value),
    \item  underlying scale (e.g. three, four, five point scale),
    \item level of expertise of the evaluators (e.g. radiologist, radiographer, scientist),
    \item number of evaluators.
\end{itemize}

\subsection{Motion Detection and Estimation}
Motion evaluation strategies can be applied if motion is \textit{explicitly} modeled within the reconstruction framework. Models detecting motion in a line-wise manner resemble classification tasks. Therefore, they can be evaluated with any classification evaluation metric as long as a ground truth exists. For an overview of classification metrics and their definitions we refer the reader to \cite{Hossin_2015}. Observed metrics specifically applied to MR motion detection tasks include accuracy, sensitivity, precision, recall, F-score and area under ROC curve \cite{Cui_2023, Zhao_2021, Oksuz_2021}. 

Motion estimates can be evaluated reference-based or reference-free. Several methods generate a reference motion field by simulating motion or obtaining the motion field through a distinct registration method. Consecutively, predicted motion parameters are compared using MAE \cite{Hossbach_2022} or RSME, which is also termed end-point-error (EPE) when applied to motion fields \cite{Qi_2021, Qi_2021b, Kustner_2022}. A further metric specifically applicable to motion fields is the end-angulation-error (EAE) \cite{Kustner_2022}, which computes the angle between the ground truth and predicted motion vector. When no motion estimate is available as reference, predicted motion fields can be used to warp images. Similar to registration evaluation, motion field accuracy can be evaluated by comparing the spatially transformed image with the target image. Any image-based similarity metric can be leveraged, whereas SSIM and PSNR \cite{Seegoolam_2019, Yang_2022}, Normalized Cross Correlation \cite{Qi_2021}, Dice Score and Hausdorff Distance on available segmentations \cite{Huang_2021, Shao_2022} have been observed in the reviewed papers. A further image-based motion evaluation strategy compares the dynamic position of relevant organ boundaries, e.g. the hepatic dome, in the predicted motion-aware reconstruction with a motion-resolved reference \cite{Terpstra_2022}.

\subsection{Downstream Tasks}
In some cases the MoCo framework does not solely aim to provide a high-quality reconstruction, but enables further downstream tasks. In this case the downstream findings can be evaluated independently, e.g. by calculating the Dice overlap on organ segmentations \cite{Oksuz_2020,Oksuz_2021, Liu_2021} or computing SSIM and relative error metrics on T2* maps \cite{Xu_2022}. 
To evaluate the added statistical power due to MoCo in longitudinal analyses, manual quality control of structural elements like cortical surface reconstructions and cortical thickness correlation analyses can be employed \cite{Duffy_2021}.
Especially if no reference is available, the sharpness of small anatomical features, such as coronary vessels can be analyzed \cite{Qi_2021, Qi_2021b}. In cardiac imaging, cardiac function analysis \cite{Ghodrati_2021, Ghodrati_2021b, Kustner_2020} or myocardial strain measurements \cite{Abdi_2023} can be evaluated. Further, as an important end goal for MR MoCo, a comparison of clinical findings in a motion corrupted and corrected scan can be conducted \cite{Liu_2020}.

\section{Discussion} \label{sec:discussion}
In the previous sections, we gave an overview of available data and motion simulation for MR MoCo reconstruction (Sec. \ref{sec:data}), we outlined the state-of-the-art model architectures for learning-based MR MoCo (Sec. \ref{sec:arch}) and their common evaluation strategies (Sec. \ref{sec:eval}). In the following, we critically discuss the reviewed methods presented in Secs. \ref{sec:data}-\ref{sec:eval}
We highlight common strategies and differences, pointing out their advantages, limitations and needs for improvement. 

\subsection{Data Availability and Motion Simulation}
\label{sec:disc_data}
The presented \textit{motion simulation} strategies (Sec. \ref{sec:data}) are predominantly used for training and evaluating motion correction approaches. Only a few motion compensation approaches include non-rigid motion simulation procedures and if so, only for evaluation \cite{Qi_2021, Qi_2021b, Zou_2022, Lyu_2021}. Non-rigid motion simulation, though, is limited both in image space \cite{Oksuz_2019, Oksuz_2020, Lyu_2021, Qi_2021, Qi_2021b, Zou_2022, Feinler_2023} and even more, the simplified version in k-space \cite{Tamada_2020, Wu_2021, Ghodrati_2021, Abdi_2023}. Simulating translation for breathing motion may broadly cover the direction of the motion but does not represent the deformable nature of real patient motion. Thus, when 3D data are available, simulation using motion fields is the more realistic and preferable approach.

Even though the motion simulation procedure based on \eqref{eq:motion_forward_model} appears to be well-defined, several variations have been implemented for both rigid-body and non-rigid motion simulation.
Some methods only simulate in-plane motion \cite{Chatterjee_2020,Lee_2021,Sommer_2020,Hossbach_2022,Oh_2021,Rotman_2021,Singh_2022,Kuzmina_2022,Cui_2023,Levac_2022,Singh_2023, Wu_2021, Tamada_2020}, whereas others consider full through-plane motion patterns, which better resemble real patient motion  \cite{Duffy_2021,Johnson_2019,Liu_2021,Pawar_2020,Pirkl_2022,Xu_2022,Haskell_2019,AlMasni_2023,Armanious_2020,AlMasni_2022,Bao_2022,Wang_2020,Eichhorn_2023,Abdi_2023}. 

Moreover, in the case of rigid-body motion, some methods limit their simulations to a small number of motion-states \cite{Chatterjee_2020,Duffy_2021,Johnson_2019, Lee_2021, Liu_2020,Usman_2020,Xu_2022,Oh_2021,Rotman_2021,Singh_2022,AlMasni_2023,Armanious_2020,Levac_2022,Singh_2023,AlMasni_2022,Wang_2020,vanderGoten_2022},
whereas others base their simulations on time-resolved motion curves, which are either generated randomly or based on clinical measurements like functional MRI time series \cite{Pawar_2020,Pirkl_2022,Sommer_2020,Haskell_2019,Hossbach_2022,Kuzmina_2022,Cui_2023,Bao_2022,Eichhorn_2023, Feinler_2023,Abdi_2023}. 
In the case of non-rigid motion, image-based simulation can only simulate discrete motion states, since multiple reconstructions are needed for merging into a corrupted image. The acquisition time for a single reconstruction limits the total number of motion states that can be simulated even for fast sequences, like cardiac cine imaging \cite{Lyu_2021} and even more, for time-consuming 3D acquisitions in the abdomen.

As described in section \ref{sec:backgr}, second-order motion-effects influence MR signal properties in addition to the effects of positional changes. A few recent approaches consider such second-order motion effects for more realistic motion simulations in specific MR sequences, like phase shifts of stimulated echoes due to respiration \cite{Abdi_2023} or motion-induced magnetic field inhomogeneity changes in T$_2$*-weighted MRI \cite{Eichhorn_2023}.

As a simplified approach to simulate motion, a few authors explicitly exclude central k-space lines \cite{Liu_2020,Oh_2021,Cui_2023,Duffy_2021, Tamada_2020, Wu_2021}, which severely limits the generalisability of their methods to real motion patterns.

In contrast to the above discussed simulation procedures for MoCo, there is in general no GT motion-free image for motion compensation methods, since breathing and especially heartbeat cannot be avoided. Breath hold and gated acquisitions have the potential to approximate GT motion-free images. Hence, motion-corrupted images can be simulated using deformation fields \cite{Qi_2021b,Qi_2021}, which can be derived from classical motion-resolved reconstructions \cite{Johansson_2018},
other imaging modalities or physical models, as e.g. the XCAT phantom \cite{Segars_2010}. 
However, such simulations might require expensive acquisitions of additional data, might not offer sufficient temporal resolution for all applications and the XCAT phantom, specifically, simulates images based on CT images, making raw k-space data unavailable. Furthermorehi w, given that the majority of presented motion compensation methods aims for acceleration, most methods focus on \textit{simulating undersampling artefacts} and compare reconstructions to fully sampled acquisitions to show the acceleration potential of their approach. These simulations vary with regard to the underlying retrospective sampling trajectory (e.g. cartesian, radial or spiral). For cartesian sampling, the center of k-space is usually explicitly sampled more frequently than the periphery, which is also common for classical acceleration methods. 

Regardless of whether motion or undersampling artefacts are simulated, many authors who train their models on simulated data at least show qualitative evaluations on a few unpaired real motion or prospectively undersampled datasets, respectively. However, some approaches only evaluate on simulated data \cite{Chatterjee_2020,Oksuz_2021,Usman_2020,Singh_2022,Kuzmina_2022,Cui_2023,Levac_2022,Singh_2023,Wang_2020,Eichhorn_2023, Wu_2021, vanderGoten_2022, Feinler_2023, Ghodrati_2021b, Lyu_2021, Wu_2021, Hammernik_2021, Pan_2022, Qi_2021b, Huang_2021, Seegoolam_2019, Yang_2022, Shao_2022, Wang_2020, Terpstra_2022, Schlemper_2018, Huang_2022, Catalan_2023, Qin_2019}, which questions their generalisability to real motion or prospectively undersampled data. 

In addition to these limitations of implemented simulation procedures, the majority of image-based approaches \cite{Chatterjee_2020,Duffy_2021,Kustner_2019,Lee_2021,Liu_2020,Oksuz_2021,Pawar_2020,Sommer_2020,Usman_2020,Tamada_2020,Lyu_2021,Jafari_2022,AlMasni_2022,Bao_2022,AlMasni_2022,vanderGoten_2022} and some k-space-based approaches \cite{Oksuz_2019,Oksuz_2020,Singh_2022,Cui_2023,Gan_2022,Shao_2022,Wang_2020} rely on magnitude-only data, which is questionable especially for the latter category. Shimron \emph{et al.}~\cite{Shimron_2022b} describe the generation of k-space data from real-valued, coil-combined magnitude data as a data crime, for instance, if zero-padding was performed during the initial reconstruction process. Furthermore, phase information is relevant for several applications, e.g. quantitative susceptibility mapping or fat-water separation, and is part of postprocessing pipelines, e.g. for background field correction. Magnitude-only approaches cannot be applied to these applications in a straightforward manner.

Public raw multi-coil datasets with paired motion experiments could enable a more realistic method development and evaluation for researchers who do not have the possibilities to acquire such raw k-space data in a paired setting. For brain imaging, currently only magnitude data with and without intentional subject motion \cite{Ganz_2022,Narai_2022} and motion-free k-space datasets \cite{Zbontar_2019} are available. For motion compensation in cardiac imaging, breath-hold cardiac gated radial dataset \cite{ElRewaidy_2020} as well as an undersampled, free-breathing k-space dataset \cite{Chen_2020} are available.

\subsection{Architectures}
A wide variety of architectures has been proposed to target MR MoCo (Sec. \ref{sec:arch}). While aiming for different applications, we outline common trends regarding the (a) data domain, (b) targeted motion types and (c) motion modeling. Furthermore, we discuss (d) model interpretability, 
(e) patient-specific models and (f) the interchangeability of modules.

\paragraph{Image-Based vs. k-Space-Based Motion Correction}
\label{arch:image_vs_kspace}

The review of proposed architectures for motion-corrected MR reconstruction shows that both image and k-space methods are commonly applied. Image-based methods profit from broadly available data, since they can be applied on existing MR image databases.
Additionally, data are frequently limited to magnitude values, which reduces the complexity of the architecture. 
Nevertheless, image-based methods are more likely to produce hallucinations, as lack of raw k-space data restricts the ability to perform data consistency checks. Also, these approaches lack the flexibility to adapt the reconstruction strategy based on motion parameters \cite{Singh_2023}.

In contrast, k-space-based methods benefit from a more comprehensive data representation that includes additional information such as phase and coil sensitivities, which can improve final image quality \cite{Lin_2021}. 
Besides a low availability of raw k-space data in practice, a potential disadvantage is that the reconstruction parameters and hardware may have a larger influence on the final image. Thus, it may be more difficult to compare results across different systems \cite{Lin_2021}.

\paragraph{Different Motion Types} 
As pointed out in Sec. \ref{sec:data}, the types of motion observed in brain and cardiac/abdominal imaging are distinct. Motion artefacts in brain images mostly originate from rigid-body motion of different severity at random time points, generally resulting in blurring. Quasi-periodic motion, which is typical in cardiac and abdominal imaging, can additionally result in ghosting artefacts \cite{Godenschweger_2016}. Because of these distinctive visual characteristics, architectures are frequently trained and tested on specific body regions. 

For \textit{image-based methods}, the presented approaches predominantly target motion correction in the brain, potentially due to the inherent capability of CNNs and encoder-decoder structures to sharpen edges, i.e. denoise artefacts apparent as blurring. To counteract periodic signal modulations leading to ghosting artefacts, abdominal and cardiac image-based methods reconstruct images from specific time intervals, i.e. motion states. Still, residual blurring persists due to continuous motion within this time interval, and missing data lead to undersampling artefacts. Therefore, image-based MoCo strategies applied to quasi-periodic moving organs mainly rely on information fusion from multiple dynamics.
Only few methods aim to learn a latent motion-free representation of the heart or abdomen from single reconstructions. 

Also, for \textit{k-space-based methods}, distinct architecture approaches exist for different motion types. The modeling of rigid-body motion with only few parameters facilitates joint optimization of motion estimation and reconstruction with a reasonable computational overhead, making it a technique typically limited to the head region (\ref{par:repl_model_based}).
With the random timing of motion in the head, some motion correction strategies focus on identifying the time of occurrence, i.e. motion detection. 
In contrast, methods targeting periodic motion mostly rely on fusion of data from different motion states, i.e. motion compensation. Since data consistency is crucial to ensure physical plausibility, considerably more k-space domain approaches have been presented than pure image-based methods for motion compensation. Whereas quasi-periodic motion compensation is the aim of most methods for the cardiac and abdominal anatomy, few methods target explicit motion correction of irregular motion sources, e.g. due to mistriggering artefacts \cite{Oksuz_2019,Oksuz_2020}.

\paragraph{Motion Modelling}
Many methods combine MoCo and reconstruction in one process (Sec. \ref{sec:k_space}). The majority of these hybrid approaches furthermore includes an explicit motion model, regardless of the optimization process or type of motion. However, such an explicit motion model only offers an approximation of the actual motion, which is considered in just one published work \cite{Feinler_2023}. As a result, the accuracy of the reconstruction is constrained \cite{Hammernik_2022b}. Nevertheless, compared to implicit motion modeling methods, explicit motion models allow for additional quality control (refer to Sec. \ref{sec:eval}).

Both explicit and implicit motion compensation techniques frequently rely on data that have been temporally separated into several motion states throughout the cardiac or respiratory cycle. Explicit methods estimate motion between the binned reconstructions, whereas implicit methods do not directly model motion but exploit temporal redundancies in the dynamic data. 
On the one hand, this requires a reliable navigator signal representing the actual motion of the organ of interest. On the other hand, binning of data from multiple cycles is susceptible to inter-cycle variability \cite{McClelland_2013}. Whereas preliminary work on motion estimation uncertainty exists \cite{Huttinga_2022b}, almost none of the presented MoCo architectures consider uncertainty in their motion modelling within the reconstruction pipeline. 

As an additional drawback of methods modelling motion based on binned motion states, the residual motion within these states blurrs the reconstruction. Increasing the temporal resolution by binning fewer data points to one motion state would lead to increased undersampling, and, therefore, affect the potential to generate reliable motion estimates. Although motion occurs continuously, many models are restricted to discrete representations. An initial attempt to avoid motion states is proposed in \cite{Huang_2022}, where a continuous representation of the motion dimension is learned. 

Lastly, current motion models integrated into learning-based MoCo architectures mainly focus on primary motion effects in k-space. Especially when handling raw data, further physics-based motion-induced secondary effects should be considered. Consequently, the motion-aware reconstruction could be extended beyond the physical motion modelling, e.g. by correcting spin history effects or B$_0$- and B$_1$-distributions~\cite{Plumley_2022}.

\paragraph{Interpretability}
When using any learning-based motion-correcting reconstruction framework, it is important to understand the model's behaviour. Employing models without such knowledge may lead to undesired effects like hallucinations, directly influencing the critical process of medical diagnosis. To avoid adverse effects of "black-box" models, interpretability should already be considered in the architecture design. Including physical knowledge, e.g. by explicit motion modeling, can aid in generating interpretable reconstruction results as well as influences of intermediate steps. For implicit models, in contrast, it is important to understand the bottle-necks. Disentangling a learned low-dimensional representation, i.e. the latent space, \cite{Liu_2021} is a first step towards such informed modeling. 

\paragraph{Patient-Specific Models}
Recently, patient-specific generative models have been developed as novel direction for motion-compensated MR reconstruction (see Sec. \ref{sec:arch_k_pure_dl}). Since motion patterns can strongly vary between patients, such individually learned representations may perform better than generalized approaches. Nevertheless, the need to retrain the model comes with prolonged reconstruction times and increased computational resources. Transferable concepts
have not yet been proposed for patient-specific generative models.

\paragraph{Interchangeable Modules}
The presented architectures for motion-corrected MR reconstruction aim at different motion patterns, anatomical regions and sequences. While a general solution is unlikely, some components can be seen as interchangeable modules to enable further development and improvement of methods. For example, varying methods for undersampled reconstruction \cite{Hammernik_2023} may be integrated into approaches that aim at removing motion-corrupted lines. Another exchangeable module can be motion estimation, which, in theory, could be conducted with any other learning-based registration method \cite{Haskins_2020}, but needs to consider strong undersampling artefacts in the input images.

While the reconstruction concept dominantly used in fetal motion correction is fundamentally different (Sec.~\ref{sec:backgr}), individual motion estimation concepts may be transferable as well. SVR-based MoCo methods frequently model rigid intra-slice motion estimation as well, e.g. based on gated recurrent units \cite{Shi_2023} or as learnable parameter within the SVR reconstruction problem \cite{Xu_2023}. Non-rigid estimation techniques or architecture backbones could be transferred from and to other applications.

\subsection{Training Objectives}
A variety of training strategies and objective functions have been adopted for optimizing MoCo models, as described in Sec. \ref{sec:obj}. Next to back-propagating errors from the motion-corrected image and from intermediate steps, like e.g. motion estimation, MoCo models can also be trained in an end-to-end fashion with a downstream task. For instance, Xu \emph{et al.} \cite{Xu_2022} combine MoCo with T$_2$* parameter quantification and Oksuz \emph{et al.} \cite{Oksuz_2020} with cardiac segmentation. For such a joint optimization, the MoCo task and the downstream task of interest might benefit from each other, improving the overall performance. However, the resulting motion corrected images might not be suitable for different downstream tasks.

Additionally, due to the limited availability of GT data (see Sec. \ref{sec:data} for details), more and more self-supervised and unsupervised methods have been proposed for data-efficient training. Adversarial training is employed to cope with unpaired image data. If k-space data are available, application of data consistency allows for self-supervised training. Proposed subject-specific generative models \cite{Huang_2022, Feng_2022, Zou_2022, Yoo_2021} are optimized by comparing the reconstruction result with measured data, and therefore, are inherently self-supervised. The Noise2Noise concept, originally proposed for image restoration \cite{Lehtinen_2018}, is adapted in a self-supervised generative \cite{Miller_2023} as well as unsupervised inter-subject \cite{Gan_2022} motion-corrected reconstruction strategy. This highlights the potential to transfer further computer vision training strategies to cope with limited data.

In general, many approaches combine various objective functions in order to guide the optimization, such as the combination with downstream losses or the combination of motion estimation or adversarial losses with image-based losses of the final reconstruction. While this can enforce specific properties in the result, like e.g. imposing more realistic motion patterns by regularizing motion fields \cite{Hammernik_2021,Qi_2021,Qi_2021b}, the training process might become more complex, since the weighting of different losses is not straightforward, but rather another hyper-parameter to be tuned. Furthermore, computational effort might increase, e.g. when combining losses in image- and k-space domain for non-Cartesian sampling patterns requiring a costly domain transformation \cite{Feng_2022}.

\subsection{Evaluation Metrics}
As outlined in Sec. \ref{sec:eval}, the most commonly used evaluation metrics are image quality metrics, which evaluate the main goal of MoCo: a high quality image. Among these, especially the full-reference metrics SSIM and PSNR stand out, which also seem to correlate well with radiological assessment \cite{Eichhorn_2022}. A downside of these full-reference methods is that they rely on the availability of paired GT data (compare Sec. \ref{sec:data}). Reference-free methods, on the other hand, do not require a GT image but are not yet widely used, since they are less consistent. 
Another important consideration for full-reference metrics is that the evaluated and GT image might not be perfectly aligned \cite{Reguig_2022}. In order to not overestimate motion-induced errors, some authors include a co-registration step before calculating full-reference metrics. However, since registration might also introduce interpolation errors, further research is needed on this topic.

In general, there is no standardized way of evaluating image quality in practice, which is not only problematic for learning-based MoCo, but extends to the entire fields of MoCo and image reconstruction. A variety of metrics is used by different authors and for some metrics, e.g. SSIM, hyperparameters can be set manually. This heterogeneity limits the comparability of different methods, even when ignoring the fact that different methods are evaluated on different datasets. Furthermore, the lack of standardized recommendations for evaluation also leaves room for "metric picking", which might lead to overestimated performances and misguide future research. 
However, when aiming to develop general recommendations, investigations on the relevance of different image quality metrics on diverse datasets are urgently needed. Since no single image quality metric can be expected to be sensitive to all possible image artefacts, such recommendations may comprise a broad, generally accepted set of metrics.

Similarly, for subjective image quality scoring, the variability of strategies regarding instructions, scales and evaluators limits the comparability of different methods. A common recommendation is to reduce inter-observer variability by averaging the scores of multiple observers. 
However, quality assessment is a time consuming process and experts such as radiologists already have a high workload in many hospitals, which limits the practicality of qualitative image evaluation. A possible solution might be to utilize deep learning models that can be trained to perform reference-free image quality assessment \cite{Kustner_2018,Esses_2018,Sujit_2019}. However, further research is needed on questions like the reliability and the generalisability of trained models to distribution shifts.

Next to the relevance of consistent image quality evaluation, we would also like to emphasize the importance of "in-between quality assurance" by evaluating motion detection and estimation as intermediate results for methods that explicitly model motion. If the extracted motion information is incorrect, these errors might propagate into the final reconstruction. Again, standardized evaluation criteria would allow for better comparisons of different methods.

Furthermore, we would like to highlight that the additional analysis of downstream tasks is application specific, which limits the potential of general recommendations. Such additional evaluations, though, might be highly relevant for the translation of developed methods into clinical practice.

\section{Conclusion and Outlook}
\label{sec:outlook}
In this review, we have provided a comprehensive overview of existing learning-based methods for MR MoCo, identifying synergies and differences in underlying data usage, architectures and evaluation strategies. In the following, we point out key findings and highlight aspects that require further investigation. 

For learning-based MoCo in MR both real and simulated data can be used for training and evaluation. Motion simulation provides the benefit of an existing ground truth and can be an effective means for initial development, however pitfalls such as sole in-plane simulation, discrete motion state modelling, exclusion of central k-space lines and erroneous processing of magnitude images need to be avoided. For non-rigid motion types, motion simulation should focus on deformable motion. Nevertheless, particularly with respect to secondary motion effects and to enable a reliable transfer to clinical applications, real data needs to be employed, at least for the evaluation process. Since such data is difficult to obtain and can strongly vary from site to site, a common database with real motion artefacts is crucial for the community. 
Inclusion of raw k-space data would further advance the development of data-consistent methods and ensure method comparability independent of individual hardware settings. 

Next to openly accessible data, we would like to emphasize the need for systematic evaluation guidelines. Up to this point, methods have been evaluated with various metrics with various definitions. Initial work on the relevance and performance of metrics needs to be extended. A standardization of both, full-reference and reference-free evaluation metrics, should be aimed for.

Architecture development should strongly focus on DC based methods, which avoid hallucinations and, thus, might be easier to translate into clinical practice. Whereas targeted motion patterns will continue to affect the architecture design, the underlying motion modelling requires careful consideration in all cases, e.g. regarding simplified assumptions of discrete motion states or uncertainties of motion estimates. Training the MoCo model in an end-to-end fashion with a downstream task to back-propagate task-specific errors seems to be promising, but is limited to the specific application. Further, self-supervised training strategies can encounter expensive acquisitions of GT data. Self-supervised patient-specific models open up a new direction, but require expensive retraining for each individual reconstruction.

In general, most state-of-the-art architectures are developed for 2D data, requiring less expensive computation. However, many 2D approaches cannot correct through-plane motion, which limits their performance for real motion-corrupted data. Also, due to the slice-selective excitation, secondary motion effects like spin history effects impact 2D data more strongly than 3D data and cannot be simulated in a straightforward manner. Thus, future research should focus, whenever possible, on 3D acquisitions or otherwise, consider 3D motion information. Exploration of transferable models may enable reduced computational burden. 

In view of the rapid development of deep learning, we expect further advances in learning-based MoCo in MRI in the near future. An initial transfer of methods developed in the machine learning community has been presented in this review, but there are many more to be exploited: methodological developments with novel architectures, e.g. diffusion models \cite{Croitoru_2023}, neural implicit representations \cite{Sitzmann_2020} and transformers \cite{Shamshad_2023}, can be advanced. Motion modelling may benefit from parallel developments of probabilistic models that include uncertainty estimates. Data-efficient strategies could reduce the need for large training data sets or long training of patient-specific models. 

Modern learning-based MoCo methods should consider the multi-dimensional nature of MRI. Clinical protocols often include multiple contrasts and dynamics, providing additional information at hand. Co-development of acquisition and motion monitoring techniques should be promoted. The presented and any future learning-based MoCo models may have the potential to be integrated into clinical MRI protocols, allowing for motion detection and estimation at fast inference times. 

With this review we aim to bridge the gap between machine learning and MRI. We see potential for further development of clinically relevant MoCo methods. Not only would this development aid in improving current clinical protocols, but open doors to areas where motion has been a major restricting factor, e.g. due to irregular and deformable patterns. Next to MR reconstruction, advances could be transferred to multi-modal imaging techniques and foster MRI as a non-invasive motion monitoring technique for applications such as PET-MR and MR-guided radiotherapy.


\printbibliography

\end{document}